\documentclass[a4paper,11pt]{article}
\usepackage{pos}

\usepackage{bbold, bm}
\usepackage{datetime}
\usepackage{numprint}
\usepackage{enumitem}
\usepackage{graphicx,color,amsmath}
\usepackage{hyperref}
\usepackage{mathtools}
\usepackage{booktabs}
\usepackage{pifont}%

\usepackage{rotating}
\allowdisplaybreaks

\newcommand{\MeV}{\ensuremath{\mathrm{MeV}}}

\newcommand{\ttt}{\ensuremath {3\to 2}}
\newcommand{\ftt}{\ensuremath {4\to 2}}

\title{From SIMP miracles to WIMP dead ends:\\
navigating the freeze out of MeV-mass dark matter}

\author[a]{Xiaoyong Chu}

\author*[b,c]{Josef Pradler}

\affiliation[a]{International Centre for Theoretical Physics Asia-Pacific, University of Chinese Academy of Sciences, 100190 Beijing, China}

\affiliation[b]{Institute of High Energy Physics, Austrian Academy of Sciences, Dominikanerbastei 16, 1010 Vienna, Austria}

\affiliation[c]{University of Vienna, Faculty of Physics, Boltzmanngasse 5, A-1090 Vienna, Austria}

\emailAdd{chuxiaoyong@ucas.ac.cn}
\emailAdd{josef.pradler@oeaw.ac.at}

\abstract{
We summarize here our studies~\cite{Chu:2022xuh,Chu:2023jyb,Chu:2024rrv} on two distinct scenarios for MeV-mass thermal dark matter freeze-out. First, we determine the minimal viable mass for dark matter below tens of MeV, considering annihilation into Standard Model particles, including photons, electrons, and neutrinos. Using a full three-sector abundance calculation, we track heat transfer between sectors and provide accurate thermal annihilation cross sections, particularly for velocity-dependent cases. The results identify fine-tuned regions where neutrino final states permit otherwise excluded p-wave annihilation scenarios. 
Second, we examine dark matter freeze-out in strongly interacting theories, where the relic abundance can be regulated not only through standard $3\pi \to 2\pi$ annihilation but also via bound-state formation~$X$, enabling effective two-body processes $XX \to \pi \pi$ and/or $\pi X \to \pi \pi$. Together, these studies highlight complementary pathways to thermal MeV-mass dark matter.
}

\FullConference{9th Symposium on Prospects in the Physics of Discrete Symmetries (DISCRETE2024)\\
 2–6 Dec 2024\\
Ljubljana, Slovenia\\}

\begin{document}
\maketitle

\section{Introduction}

The thermal production of dark matter (DM) provides one of the most compelling frameworks for explaining the observed DM abundance. In the simplest scenarios, DM freezes out through two-body annihilation into Standard Model (SM) particles. Here, sub-GeV thermal relics face tensions with cosmological constraints. In particular, for DM in the MeV mass range, thermal freeze-out coincides with the epochs of neutrino decoupling and electron-positron annihilation. This requires a coupled-sector treatment, to correctly predict both the relic abundance and the effective number of relativistic degrees of freedom, $N_{\rm eff}$.  A systematic approach to this problem was developed in~\cite{Chu:2022xuh,Chu:2023jyb}, where the formalism incorporates in a numerically stable way the detailed balancing of equilibrium rates and factorizes neutrino and dark matter chemical potentials within the relevant collision terms.

An alternative thermal freeze-out scenario arises when DM is only weakly coupled to the SM but depletes its number density via self-interactions, as in the strongly interacting massive particle (SIMP) mechanism. Here, the relic abundance is set by $3\to2$ or $4\to2$ DM-only interactions, rather than direct annihilation into SM states. This framework, originally proposed with odd-numbered Wess-Zumino-Witten (WZW) interactions, allows for DM to remain thermally coupled while avoiding stringent constraints from direct detection. However, the presence of bound states in the dark sector introduces new annihilation channels that can significantly modify the standard SIMP freeze-out picture, opening additional pathways for number-changing interactions and altering relic density predictions. This is the subject of the work~\cite{Chu:2024rrv}.

 \begin{figure}[tb]
\begin{center}
\includegraphics[width=0.4\textwidth,viewport=0 710 700 1100,clip=true]{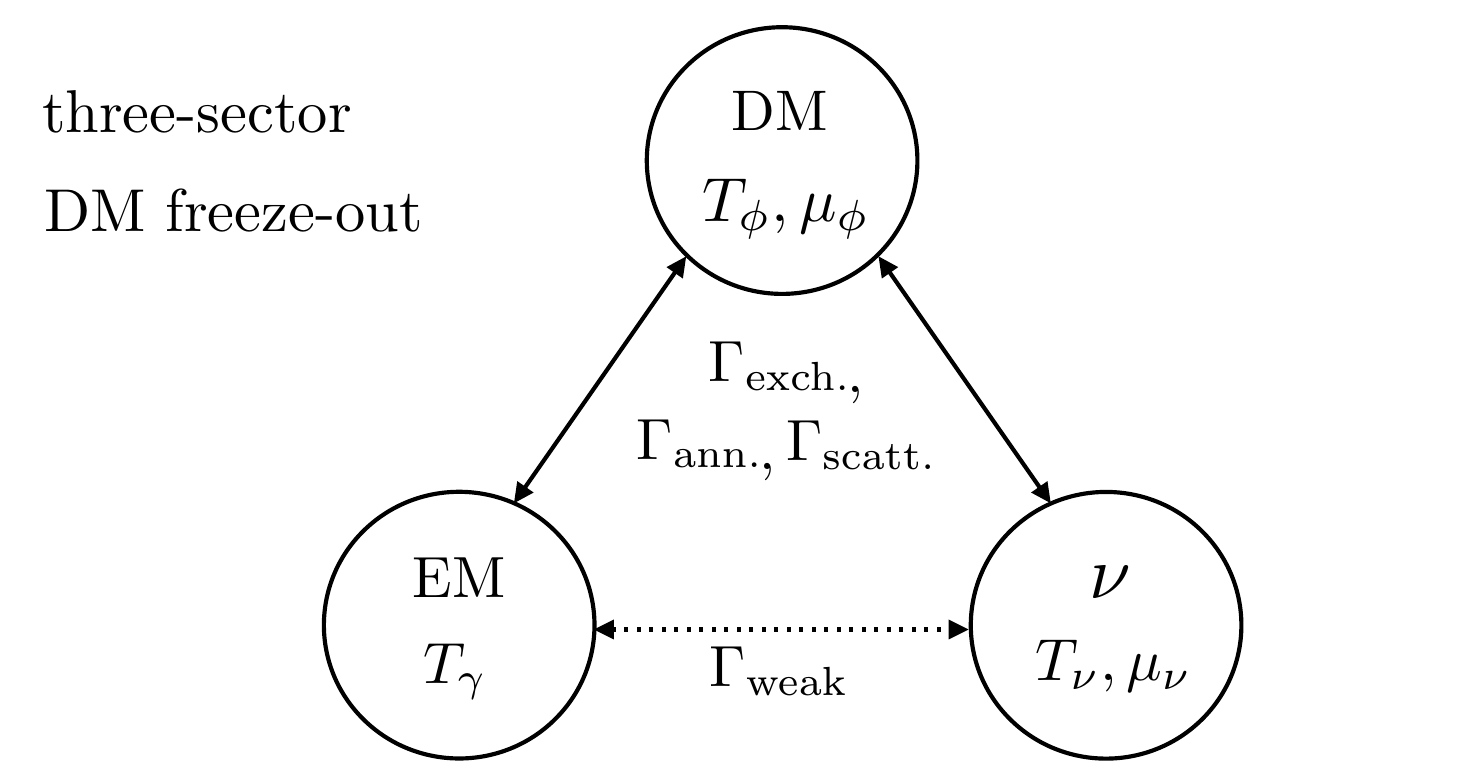}%
\includegraphics[width=0.6\textwidth]{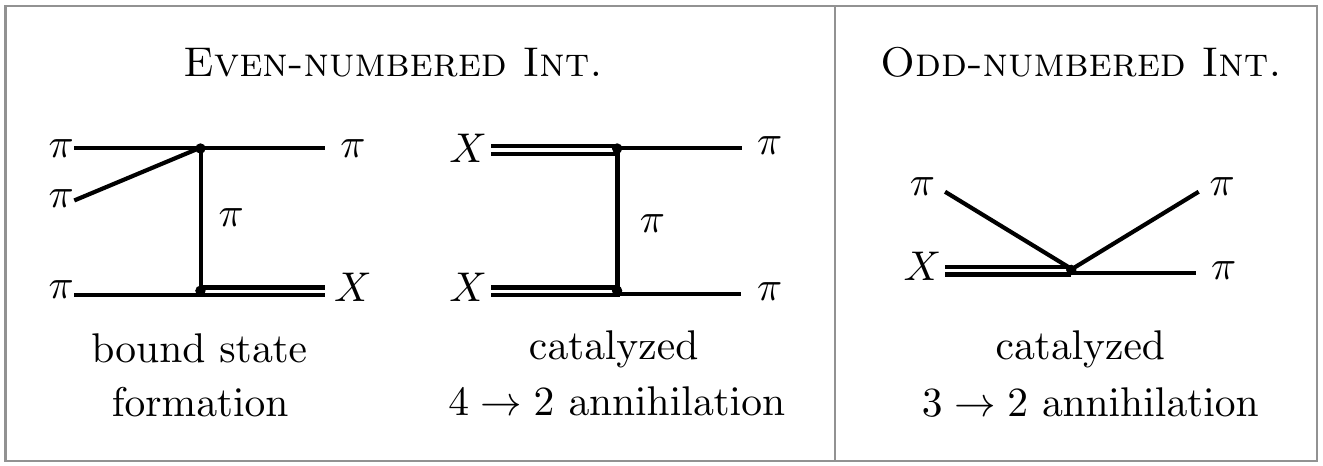}
\caption{
{\it Left:} Schematic for freeze out of MeV-mass thermal, weakly coupled light dark matter showing the interplay of the three coupled sectors~\cite{Chu:2022xuh,Chu:2023jyb}. {\it Right:} Key processes in SIMP freeze out: $X$-formation, $X$-assisted annihilation via four-point interactions, and catalyzed $3\to 2$ annihilation~\cite{Chu:2024rrv}.
}
\label{fig:scheme}
\end{center}
\end{figure}

In this presentation, we unify these perspectives by reporting on those two distinct but related thermal freeze-out mechanisms: (i) MeV-scale DM that remains in equilibrium with the SM plasma and affects cosmological observables such as $N_{\rm eff}$, and (ii) bound-state formation in SIMP models, which catalyzes annihilation processes and reshapes the dynamics of thermal freeze-out. In (i) we employ a fully self-consistent three-sector treatment to determine the precise relic density and $N_{\rm eff}$ constraints for light DM annihilating into both neutrinos and electromagnetically interacting particles. In (ii), we investigate the impact of bound-state-assisted annihilation in strongly interacting dark sectors, demonstrating that such processes can alleviate tensions between the strength of the self-scattering cross section and the required interaction strength from the relic density requirement.

\section{MeV-mass thermal light dark matter freeze out}

The three-sector system under consideration is schematically illustrated in the left panel of Fig.~\ref{fig:scheme}. Its evolution is governed by several key rates, including the weak interaction rate, $\Gamma_{\rm weak}  \equiv  n_e G_F^2 T_\gamma^2$, which determines neutrino decoupling in a standard cosmological history, and the total DM annihilation rate, $\Gamma_{\rm ann.}  \equiv n_\phi \langle \sigma_{\rm ann.} v \rangle$, which sets the DM relic abundance through the chemical decoupling from the SM plasma. Here, $G_F$ is the Fermi constant, $T_\gamma$ the photon temperature, and $n_{e/\phi}$ the number densities of electrons and DM, respectively.  Energy transfer between DM and the electromagnetic (EM) or neutrino sectors is dictated by the rate $\Gamma_{{\rm exch.},i}  \equiv n_\phi^2  \langle \sigma_{{\rm ann.}, i} v \delta E\rangle /\rho_{i}$, where the thermal average accounts for the exchanged energy $\delta E$. If $\Gamma_{\rm weak}<H$ but $\Gamma_{{\rm exch.},i} >H$ for $i= \nu$ and $e/\gamma$, neutrinos may remain thermally coupled to photons beyond standard decoupling, directly affecting $N_{\rm eff}$.  Energy exchange also occurs via elastic scatterings, described by $\Gamma_{{\rm scatt.}i} \equiv n_\phi n_i \langle \sigma_{\rm scatt.}^{\phi, i}  v \,\delta E\rangle / \rho_i$, where the exchanged energy scales with the sector temperature difference. For further details on these rates and their role in DM decoupling, we refer to~\cite{Chu:2022xuh}.

The total annihilation cross section times the M{\o}ller velocity $v_M$ follows the standard non-relativistic expansion in terms of the relative velocity $v_{\rm rel}$,%
\begin{align}
    \sigma_{\rm ann} v_{M} = a + bv_{\rm rel}^2 + \mathcal{O}(v_{\rm rel}^4).
\end{align}
A non-relativistic thermal average gives $\langle \sigma v_M \rangle = a + 6b/x + \dots$, using $\langle v_{\rm rel}^2 \rangle  \simeq  6T_\phi /m_\phi = 6/x$.  
The thermal histories of DM candidates at the MeV mass scale, considering annihilation into electromagnetically interacting particles and neutrinos with branching ratios ${\rm Br}_{\rm EM}$ and ${\rm Br}_\nu$, respectively, satisfying ${\rm Br}_{\rm EM} + {\rm Br}_\nu = 1$ is considered below.

\begin{figure*}[t]
\begin{center}
\includegraphics[width=0.485\textwidth]{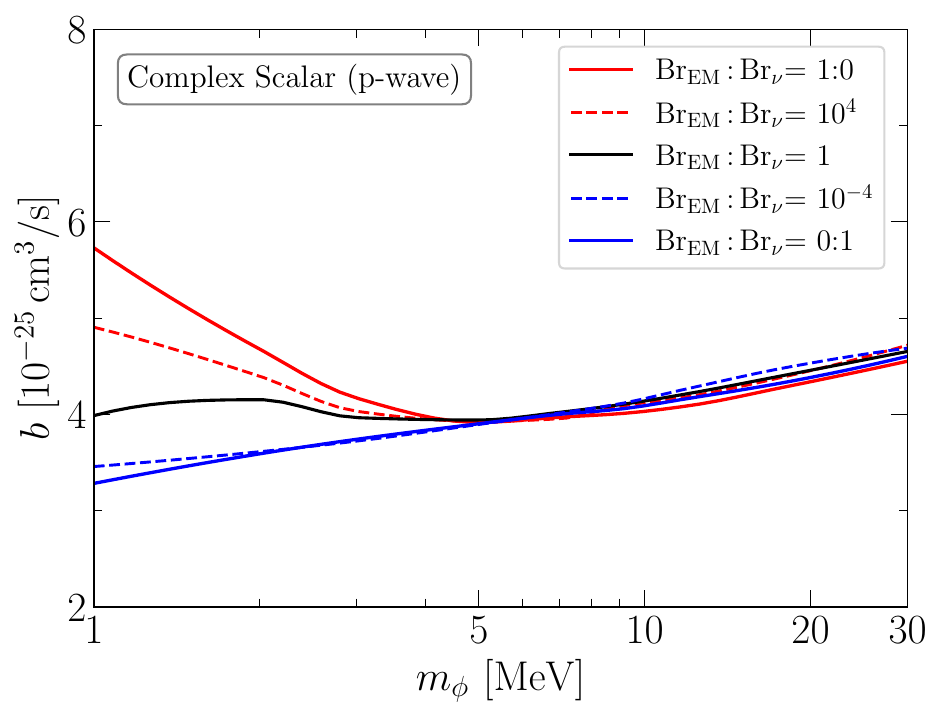}
\includegraphics[width=0.49\textwidth]{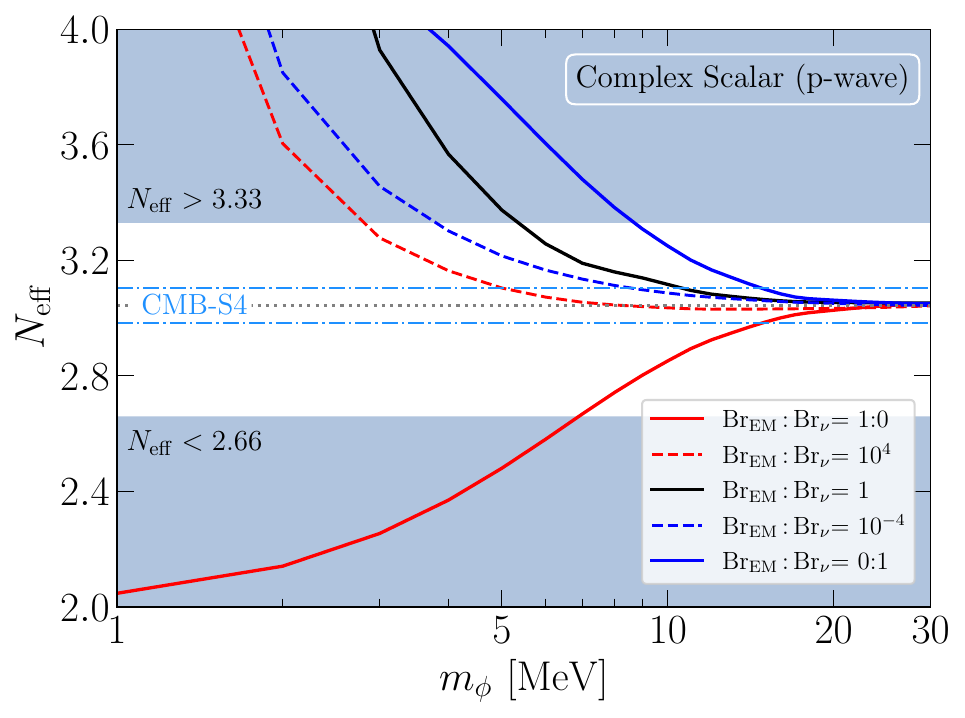}
\end{center}
\caption{{\it Left:} Thermal annihilation cross sections for a $p$-wave annihilating complex scalar $\phi$, obtained from solving the coupled three-sector system. 
{\it Right:} associated obtained $N_{\rm eff}$ values. The different colors and linestyles represent varying branching ratios.}
\label{fig:thermal_xsec}
\end{figure*}

The annihilation cross section into the EM sector is subject to strong cosmological constraints.  For $s$-wave DM in the MeV mass range, the most stringent limits come from Planck CMB observations, which impose $a \times {\rm Br}_{\rm EM} \lesssim (3\text{--}4)\times 10^{-30}\,{\rm cm}^3/{\rm s}$~\cite{Slatyer:2015jla}. Consequently, for thermal $s$-wave freeze-out, ${\rm Br}_{\rm EM}$ must be below $10^{-4}$. In contrast, due to velocity suppression, $p$-wave freeze-out with $b \times {\rm Br}_{\rm EM} \sim 10^{-25}\,{\rm cm}^3/{\rm s}$ remains unconstrained by CMB and low-redshift indirect searches~\cite{Boudaud:2016mos, Boudaud:2018oya}. Constraints on DM annihilation into neutrinos are much weaker. For DM masses below $20~\MeV$, values of $a \times {\rm Br}_{\nu} \lesssim (1\text{--}100)\times 10^{-24}\,{\rm cm}^3/{\rm s}$ remain allowed based on Borexino, KamLAND, and Super-Kamiokande data~\cite{Arguelles:2019ouk}, leaving ample parameter space for $s$-wave thermal freeze-out, and even more so for $p$-wave scenarios.  

Incorporating elastic scattering between DM and electrons/neutrinos requires a specific model, as no universal relation exists between annihilation and scattering rates. The models we use are detailed in~\cite{Chu:2022xuh} and~\cite{Chu:2023jyb}. 
The induced model dependence, however, is relatively mild. For $s$-wave annihilation, elastic scattering is irrelevant, and the model dependence cancels out. For $p$-wave annihilation, we assume a minimal scenario where the same heavy mediator responsible for annihilation also governs elastic scattering, ensuring broad applicability.%

Solving the full set of Boltzmann equations determines the thermal annihilation cross sections required for $\phi$ in the coupled three-sector system, as shown in Fig.~\ref{fig:thermal_xsec}. The left panel corresponds to $s$-wave annihilation. For $m_{\phi} \ge 25\,$MeV, DM freeze-out occurs well before neutrino-electron decoupling, so the branching ratio has little impact.  For $m_{\phi} \lesssim 10\,$MeV, $s$-wave freeze-out coincides with the EM sector being reheated by electron-positron annihilation. This elevates $T_\gamma$ relative to $T_\nu$, and DM coupled primarily to the EM sector remains more abundant compared to DM interacting mainly with neutrinos. To achieve the correct relic density, DM with dominant electron coupling must have a larger annihilation cross section.

For intermediate branching ratios, the dynamics become more complex, depending on energy transfer between the three sectors. The dark sector can maintain kinetic equilibrium between the EM and neutrino sectors (\textit{i.e.}, $T_\nu = T_\gamma$) even after neutrino decoupling, thereby increasing $T_\nu/T_\gamma$ relative to standard cosmology. Additionally, DM annihilation after EM-neutrino kinetic decoupling can further modify $T_\nu/T_\gamma$, depending on $\text{Br}_{\rm EM}:\text{Br}_\nu$. In most cases, the first effect dominates. Notably, by tuning $\text{Br}_{\rm EM}:\text{Br}_\nu$, one can balance these effects, recovering the standard cosmological value $T_\nu/T_\gamma \approx 0.7164$. For $s$-wave annihilation, the dominant energy transfer occurs via DM pair creation/annihilation, with neutrino-EM kinetic decoupling primarily sensitive to $\text{Br}_\nu\,\text{Br}_{\rm EM}$.

The results for $N_{\rm eff}$ using the canonical DM annihilation cross section are shown in the right panel of Fig.~\ref{fig:thermal_xsec}; 
 the standard cosmological value is  $N_{\rm eff}^{\rm SM} = 3.044$ and current Planck and BAO data constrain it to $2.66 \leq N_{\rm eff} \leq 3.33$ at 95\% C.L.~\cite{Planck:2018vyg}. The Simons Observatory aims for a sensitivity of $|\Delta N_{\rm eff}| \lesssim 0.1$~\cite{SimonsObservatory:2018koc}, while CMB-S4 is projected to reach $|\Delta N_{\rm eff}| \lesssim 0.06$~\cite{Abazajian:2019eic}, assuming standard cosmological history.  
In most scenarios, energy transfer from the EM sector to neutrinos via the dark sector enhances $N_{\rm eff}$. The effect is primarily governed by $\text{Br}_{\rm EM} \,\text{Br}_\nu$, meaning that cases with $\text{Br}_{\rm EM}:\text{Br}_\nu = 10^{4}$ and $10^{-4}$ yield similar outcomes.

\begin{figure}[t]
\begin{center}
\includegraphics[width=0.5\textwidth]{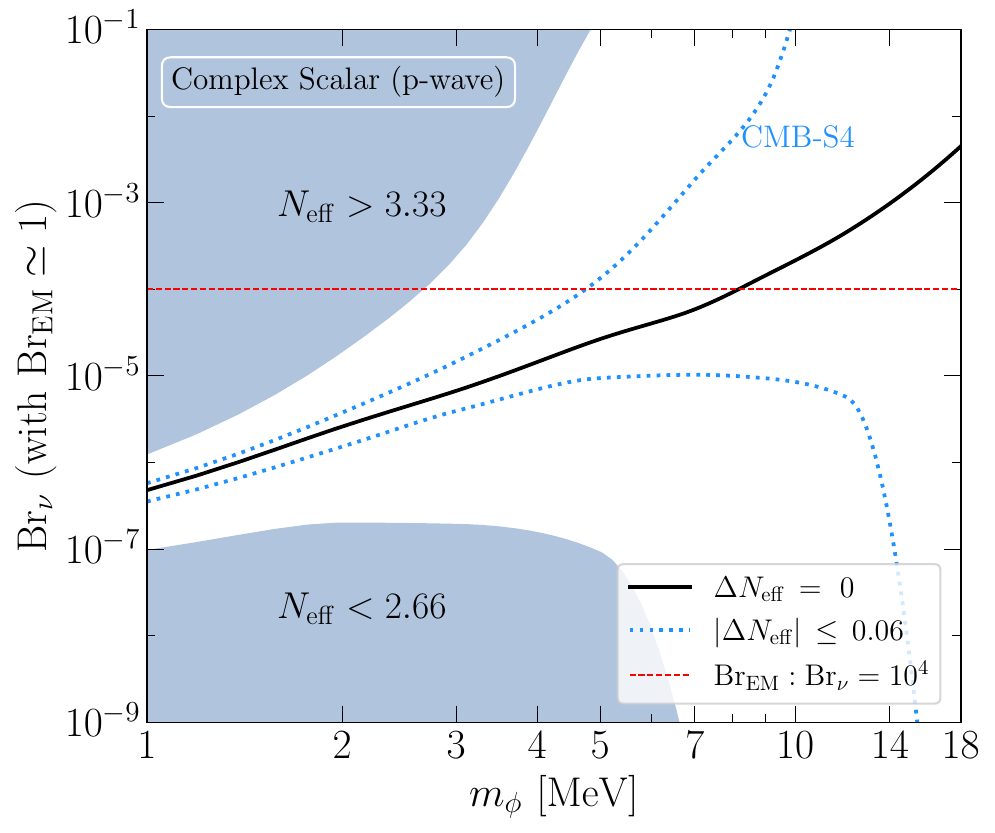}
\caption{Finely-tuned branching ratio that yield diminishing $N_{\rm eff}$  at recombination. The increase of $N_{\rm eff}$ due to delayed kinetic decoupling of  EM and neutrino sectors is compensated by the subsequent heating of the EM sector by DM annihilation post kinetic decoupling.  A too high primordial helium abundance excludes points along the black line for $m_\phi < 3\,\MeV$.
}
\label{fig:SPoptimal}
\end{center}
\end{figure}

The dark sector influences $N_{\rm eff}$ through two competing mechanisms. First, DM interactions can prolong EM-$\nu$ kinetic equilibrium, which invariably increases $N_{\rm eff}$ relative to its standard value.  
This occurs because, in conventional cosmology, $T_\nu < T_\gamma$ after $e^\pm$ annihilation. If DM maintains thermal contact between the EM and neutrino sectors for an extended period, neutrinos receive additional energy, raising their temperature relative to photons. 
Second, DM annihilation \textit{after} EM-$\nu$ kinetic decoupling can either raise or lower $N_{\rm eff}$, depending on the relative branching fractions. If $\text{Br}_{\rm EM} \gg \text{Br}_\nu$, more energy is deposited into the EM sector, reducing $N_{\rm eff}$. When $\text{Br}_{\rm EM} : \text{Br}_\nu \gg 1$, these two effects can cancel each other.  
For this cancellation to occur, the kinetic decoupling must happen well before the DM abundance undergoes significant Boltzmann suppression, requiring $\text{Br}_{\rm EM} \text{Br}_\nu \ll 1$. 

This cancellation is illustrated by the solid black curve in Fig.~\ref{fig:SPoptimal} for a complex scalar DM candidate with $p$-wave annihilation. The analogous scenario for $s$-wave annihilation is already ruled out by indirect DM searches. 
A similar pattern holds for Dirac DM, though a greater degree of fine-tuning is required due to its larger number of degrees of freedom.  

However, the additional DM energy density  contribution during freeze out affects the predicted primordial abundances of light elements. Inputting the evolution of the neutrino and photon temperatures, along with the DM energy density, into a BBN code~\cite{Pospelov:2010hj}, 
we observe that while the increase in (D/H) remains at or below $2\%$, the rise in the helium abundance rules out finely tuned branching ratios for $m_\chi \leq 2\ \MeV$.  

\section{Bound state assisted SIMP freeze out}

Bound states $X \equiv [\pi\pi]$ among SIMPs, denoted as $\pi$, can significantly impact relic abundance predictions, introducing new freeze-out channels that catalyze annihilation processes, in addition to bound state formation reactions, 
\begin{subequations}
\begin{align}
\label{eq:cat-wzw} \text{\textit{Catalyzed} \ttt\ annihilation: }  & \    \pi + X \to \pi +  \pi \,, 
 \\
\label{eq:cat-ftt}
\text{\textit{Catalyzed} \ftt\ annihilation: } & \  X +  X\to \pi + \pi\, , \\
\text{\textit{Guaranteed} $X$ formation: } &\  \pi +  \pi +  \pi \to   \pi + X  \,.
\label{eq:guaranteed-formation}
\end{align}
\end{subequations}
Those reactions are illustrated in the right panel of Fig.~\ref{fig:scheme}.
The first two effective $2\to2$ reactions compete with the usual $3\pi\to 2\pi$ and $4\pi\to 2\pi$ processes in depleting the overall DM density. Notably, while $3\pi\to 2\pi$ and \eqref{eq:cat-wzw} share the same underlying odd-numbered interaction, the process $X X\to \pi \pi$ can proceed purely via \emph{even-numbered} interactions, such as a four-point self-coupling.  If the theory permits sufficiently long-lived $X$ states, their formation is inevitable via the radiationless exoergic process shown in the last reaction above. This reaction can also proceed through even-numbered interactions and is not suppressed compared to the standard $3\to 2$ process. Consequently, $X$ is expected to form efficiently; see Fig.~\ref{fig:scheme} for an illustration.

Choosing a scenario that closely parallels the original works on the SIMP mechanism~\cite{Hochberg:2014dra,Hochberg:2014kqa}, we may consider a low-energy effective theory describing $a = 1, \dots, N_\pi$ massive pseudo-Goldstone bosons $\pi_a$, which arise as DM candidates from a confining dark non-Abelian gauge sector with $N_f$ fermionic degrees of freedom; many details on the concrete choice are found in~\cite{Kulkarni:2022bvh}.
We note, however, that the underlying dynamics of the proposal are more general.  
For a flavor-degenerate quark mass matrix $M$ with entries $m$, the resulting pseudo-Goldstone bosons acquire a universal mass given by  $m_\pi $, and $f_\pi$ is the decay constant. The leading interaction terms in the chiral Lagrangian take the form  
\begin{align}
\label{eq:chiralLint}
    \mathcal{L}^{\rm even}_{\rm int} & \supset  -\frac{1}{3 f_{\pi}^{2}} \operatorname{Tr}\left(
[\pi,\partial_\mu \pi][\pi,\partial^\mu \pi]   
    \right)
    +\frac{m_{\pi}^{2}}{3 f_{\pi}^{2}} \operatorname{Tr}\left[ \pi^{4} \right] + \dots, 
\end{align}
Odd-numbered interactions, represented by a non-vanishing Wess-Zumino-Witten (WZW) term, arise only if the symmetry-breaking pattern leads to a coset space with a non-trivial fifth homotopy group~\cite{Witten:1983tw}. The leading-order WZW Lagrangian is given by  
\begin{align}
\label{eq:LWZW}
    \mathcal{L}_{\rm int}^\text{odd} & =\frac{2 N_{c}}{15 \pi^{2} f_{\pi}^{5}} \epsilon^{\mu \nu \rho \sigma} \operatorname{Tr}\left[\pi \partial_{\mu} \pi \partial_{\nu} \pi \partial_{\rho} \pi \partial_{\sigma} \pi\right].
\end{align}

\begin{figure}[t]
\centering
\includegraphics[width=0.6\textwidth]{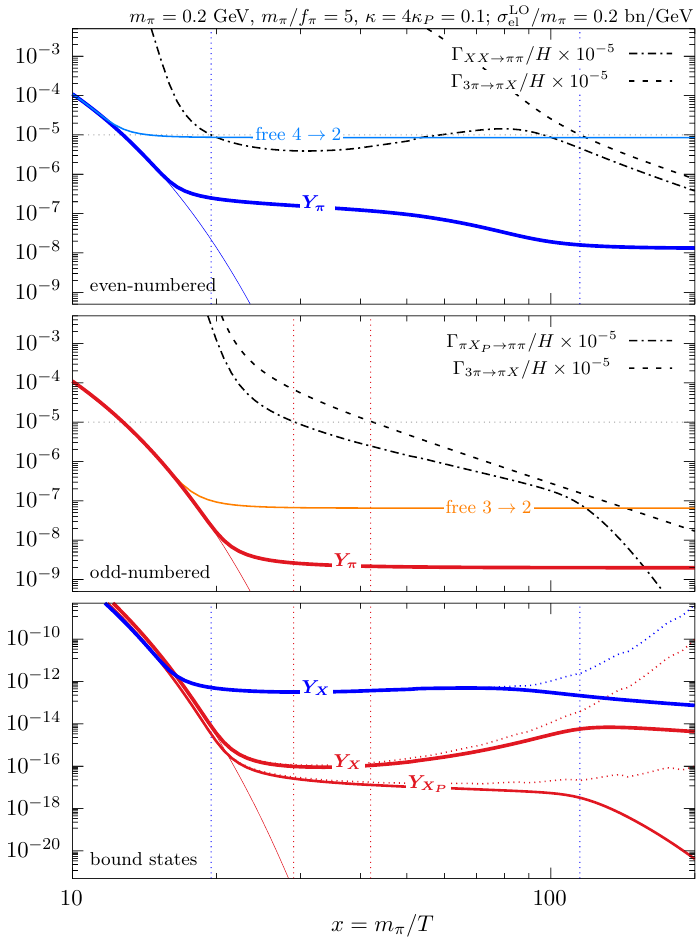}
\caption{Evolution of the DM abundance, $Y_{\pi} + 2Y_{X} \simeq Y_{\pi}$, 
for the parameter set $m_\pi = 0.2\,$GeV, $m_\pi/f_\pi = 5$, $\kappa_S = 4\kappa_P = 0.1$, $R(0) = 0.3m_\pi^{3/2}$, and $dR(0)/dr = 0.06m_\pi^{5/2}$. 
\textit{Top:} Scenario with only even-numbered interactions. The freeze-out of $XX$ annihilation occurs at $x_1 \simeq 20$, while bound-state formation ceases at $x_2 \simeq 115$. 
\textit{Middle:} Case including odd-numbered WZW interactions. The freeze-out of $X_P$ annihilation is delayed compared to the even-numbered case, allowing chemical equilibrium to persist longer. For comparison, the freeze-out of the standard $4\to2$ and $3\to2$ processes is also shown in the top and middle panels.  
\textit{Bottom:} Evolution of bound-state abundances, displayed using the same color scheme as the corresponding cases above.  
With the chosen parameters, the observed DM relic abundance is achieved in the odd-numbered scenario. For the even-numbered case, the correct abundance is obtained by adopting $R(0) = 1.4m_\pi^{3/2}$.}  
\label{fig:sol}
\end{figure}

For the bound state formation process $3\pi \to \pi X$ the dominant contributions are from $t$- and $u$-channel diagrams, as illustrated in Fig.~\ref{fig:scheme}. The propagator denominator is enhanced by a kinematic matching condition, $k^2 - m_\pi^2 \propto -E_B m_\pi$, where $E_B$ is the bound state binding energy. 
A similar approach can be taken to compute the annihilation cross section for $XX \to \pi \pi$. We restrict our focus to $S$-wave initial bound states. The process involves six $t$- and $u$-channel diagrams, which become related in the limit where the internal motion of the bound-state constituents is neglected. Additionally, when including odd-numbered $3\to2$ interactions mediated by the WZW term in~\eqref{eq:LWZW}, we obtain the corresponding cross section for the $\pi X \to \pi \pi$ process.
The parametric dependencies of the respective cross sections read 
\begin{align}
 \langle\sigma_{3\pi\to \pi X}v^2\rangle \propto 
  \frac{R_S^2(0)}{ f_\pi^8} \left( \frac{m_\pi}{E_B} \right)^{3/2} ,\quad
\label{eq:XXtopipi}
\langle\sigma_{XX\to \pi\pi} v \rangle \propto \frac{R_S^4(0) }{ f_\pi^8 } \,, 
\quad 
\langle \sigma_{\pi X \to \pi\pi} v \rangle \propto \frac{ m_\pi^3 R_P'^2(0)} {f_\pi^{10} x} .
\end{align}
The full expressions are found in the original work~\cite{Chu:2024rrv}. In the above expressions, it is assumed that the typical kinetic energy of incoming particles satisfies \( m_\pi \langle v^2\rangle/2 \lesssim E_B \), which is equivalent to the condition $T_{\rm f} \lesssim E_B$.  The first two reactions involve the $S$-wave radial bound state wave function at the origin, defined by $R(0)=\sqrt{4\pi}|\psi(0)|$. The last reaction requires $X$ to be in a $P$-wave state, $X_P$, with orbital angular momentum $\ell = 1$; $R_P'(0)$ is the radial derivative of the $P$-wave radial wave function, evaluated at the origin.

\begin{figure}[t]
\centering
\includegraphics[width=0.7\textwidth]{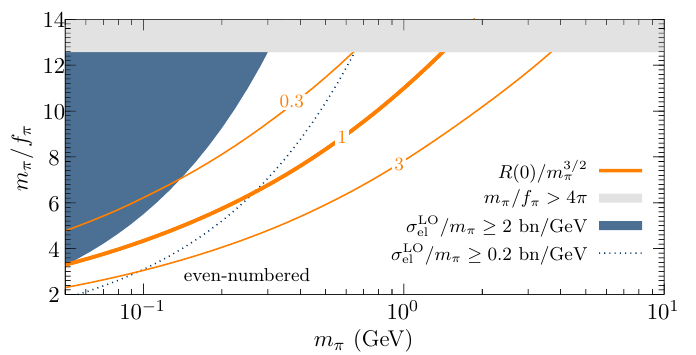}
\caption{Contours of the observed DM abundance in the even-numbered scenario for $\kappa = 0.1$ and initial conditions $R(0)/m_\pi^{3/2} = 0.3,\,1,\,3$. The shaded region (dotted line) represents the DM self-scattering constraint, requiring $\sigma^{\rm LO}_{\rm el} \geq 2$\,($0.2$)\,bn/GeV. For the free $4\pi \to 2\pi$ process to be viable at $m_\pi > 1$\,MeV, the condition $m_\pi/f_\pi > 4\pi$ would need to be satisfied.  
}
\label{fig:evenscan}
\end{figure}

The evolution equations governing the two populations, free $\pi$ and bound states $X$, are solved in terms of their comoving number densities, normalized to the total entropy density $s$, $Y_{\pi,X} = n_{\pi,X}/s$ (flavor sum implicit). Kinetic equilibrium with the SM sector is assumed.
At high temperatures $T$ (small $x = m_\pi/T$), both $\pi$ and $X$ track their equilibrium distributions. 
For the even-numbered case,
chemical decoupling occurs around $x_1 \sim 20$, when the annihilation rate satisfies  
$n_X^2  \langle \sigma_{XX\to \pi\pi} v\rangle /n_\pi \simeq H(x_{1})$.  
The subsequent evolution for $x > x_1$ is non-trivial, as bound state formation via $3\pi \leftrightarrow \pi X$ remains active. This process maintains detailed balance between the $X$ and $\pi$ populations. Eventually, bound state formation freezes out at $x_2 > x_1$, when the reaction rate satisfies  
$\Gamma_{3\pi \to \pi X} \equiv n_\pi^2 \langle \sigma_{3\pi \to \pi X} v^2 \rangle  = H(x_2)$. To within a factor of two, the solution of the abundance evolution can be expressed in suggestive form, 
\begin{align}\label{eq:BoltzSolApp}
\Omega^{\rm even}_\pi \sim  0.2
\left( \frac{200\,\kappa x_2^5}{e^{2\kappa x_2}} 
\frac{{\rm bn/GeV}}{ \langle \sigma_{XX\to \pi\pi} v \rangle/m_\pi} \frac{m_\pi}{\text{GeV}}\right)^{1/3} ,
\end{align}
where $\kappa = E_B/m_\pi$. 
We note that the relic abundance is determined by $x_2$, the freeze-out point of bound state formation, rather than $x_1$. Second, there is a strong sensitivity to $\kappa$. Given that $x_2$ typically falls between 50 and 100, a value of $\kappa = 0.1$ implies sub-GeV DM with self-interactions on the order of cm$^2$/gram—comparable to the odd-numbered SIMP scenario. Lastly, in contrast to standard $2\to2$ freeze-out, the relic density here exhibits a weaker dependence on the annihilation cross section, scaling as its cubic root.  As a function of SIMP mass, contours of correct relic density are shown in Fig.~\ref{fig:evenscan}.

In the odd-numbered scenario, chemical decoupling occurs when the interaction rate satisfies  
$n_{X_P}  \langle\sigma_{\pi X_P \rightarrow \pi\pi} v\rangle  \simeq H(x_1)$. Since the ratio of $P$-wave to $S$-wave bound states $Y_{X_P}/Y_{X_S}$—and consequently the reaction rate of $\pi X_P\to \pi\pi$—decreases exponentially for $x > x_1$, the $\pi$ population effectively freezes out at $x_1$. If bound state formation via $3\pi \to \pi X$ persists beyond this point ($x_2 > x_1$), it leads to the following expression for the relic abundance,
\begin{equation}
\label{eq:oddBoltzSol}
   \Omega^{\rm odd}_\pi \simeq  0.2 \,\left(\frac{ x_1}{20}\right)^{5/4} \left(\frac{e^{-\kappa_P x_1}\,10^{-3} \,{\rm bn/GeV}}{\langle\sigma_{\pi X_P \rightarrow \pi\pi} v\rangle/m_\pi} \right)^{1/2}\,. 
\end{equation}  
The various  panels of Fig.~\ref{fig:sol} show the full numerical freeze-out solutions for both cases for free pions and bound states alike.

As in all SIMP scenarios where freeze-out occurs via self-depletion, maintaining kinetic equilibrium with the radiation bath is essential to ensure a {\it cold} DM population.  
For this to hold, an elastic scattering process of the form $\pi\,{\rm SM}_i \to \pi\,{\rm SM}_i$ must be active, with a rate  
$\Gamma_{\pi\,{\rm SM}} = \langle \sigma_{\pi\,{\rm SM}} c \rangle n_i$  
that exceeds the Hubble expansion rate, $\Gamma_{\pi\,{\rm SM}} > H$, during freeze-out. If such a process is efficient, it also implies that $\pi\pi \to {\rm SM}_i \overline{\rm SM}_i$ annihilation can occur.  
For the SIMP mechanism to remain viable, the annihilation rate must satisfy  
$\Gamma_{\rm ann}  = n_\pi \langle  \sigma_{\rm ann} v \rangle  < H$,  
where $\sigma_{\rm ann}$ denotes the cross section for $\pi\pi \to {\rm SM}_i \overline{\rm SM}_i$. Both conditions are naturally fulfilled in most scenarios~\cite{Hochberg:2014dra}.  
Assuming a constant annihilation cross section, $\sigma_{\rm ann} v \simeq \text{const.}$, we can estimate the induced decay width of $X$ as  
$\Gamma_X \sim  |\psi(0)|^2 (\sigma_{\rm ann} v)$,  
where $\sigma_{\rm ann} v$ corresponds to the $\pi\pi \to {\rm SM}_i \overline{\rm SM}_i$ annihilation cross section.  
When contact interactions are involived, elastic and annihilation cross sections are often comparable, $\sigma_{\pi\,{\rm SM}} c \sim \sigma_{\rm ann} v$ and we may  estimate, 
\begin{align}
 1\lesssim   \frac{\Gamma_{\pi\,{\rm SM}}}{H} \lesssim 
    \frac{10^6}{x^3}
     \left(\frac{m_\pi}{100\ \MeV} \right)^3 \frac{\MeV^3}{|\psi(0)|^2} \,.
\end{align}
At freeze-out, this condition is easily satisfied for $|\psi(0)| < m_\pi^{3/2}$. We thus conclude that kinetic equilibrium can be maintained without compromising the longevity of $X$, while still ensuring sub-Hubble two-body annihilation rates. Consequently, the requirements for coupling the dark sector to the SM remain comparable to those in the standard SIMP framework.

\section{Conclusion}

These proceedings explore two distinct yet related mechanisms for thermal dark matter  freeze-out, both constrained by the same mass scale. First, we studied the impact of MeV-scale thermal DM on cosmological observables, particularly the effective number of relativistic degrees of freedom, $N_{\rm eff}$. Second, we analyzed the role of bound states in modifying the conventional SIMP framework. Despite their differences, both scenarios underscore the importance of thermal equilibrium dynamics in shaping the properties of sub-GeV DM candidates.  
Together, these studies provide a more complete picture of the viable parameter space for MeV-scale thermal DM. While simple thermal relics in this mass range have long been considered at the edge of cosmological viability, our results offer a rigorous and self-consistent framework that charts out in better detail the viable parameter spaces.

\paragraph*{Acknowledgments.}
Funded/Co-funded by the European Union (ERC, NLO-DM, 101044443). This work was also supported by the Research Network Quantum Aspects of Spacetime (TURIS).

\bibliographystyle{JHEP}
\bibliography{refs}

\providecommand{\href}[2]{#2}\begingroup\raggedright\begin{thebibliography}{10}

\bibitem{Chu:2022xuh}
X.~Chu, J.-L. Kuo, and J.~Pradler, {\it {Toward a full description of MeV dark matter decoupling: A self-consistent determination of relic abundance and Neff}},  {\em Phys. Rev. D} {\bf 106} (2022), no.~5 055022, [\href{http://arxiv.org/abs/2205.05714}{{\tt arXiv:2205.05714}}].

\bibitem{Chu:2023jyb}
X.~Chu and J.~Pradler, {\it {Minimal mass of thermal dark matter and the viability of millicharged particles affecting 21-cm cosmology}},  {\em Phys. Rev. D} {\bf 109} (2024), no.~10 103510, [\href{http://arxiv.org/abs/2310.06611}{{\tt arXiv:2310.06611}}].

\bibitem{Chu:2024rrv}
X.~Chu, M.~Nikolic, and J.~Pradler, {\it {Even SIMP miracles are possible}},  {\em Phys. Rev. Lett.} {\bf 133} (2024), no.~2 2, [\href{http://arxiv.org/abs/2401.12283}{{\tt arXiv:2401.12283}}].

\bibitem{Slatyer:2015jla}
T.~R. Slatyer, {\it {Indirect dark matter signatures in the cosmic dark ages. I. Generalizing the bound on s-wave dark matter annihilation from Planck results}},  {\em Phys. Rev. D} {\bf 93} (2016), no.~2 023527, [\href{http://arxiv.org/abs/1506.03811}{{\tt arXiv:1506.03811}}].

\bibitem{Boudaud:2016mos}
M.~Boudaud, J.~Lavalle, and P.~Salati, {\it {Novel cosmic-ray electron and positron constraints on MeV dark matter particles}},  {\em Phys. Rev. Lett.} {\bf 119} (2017), no.~2 021103, [\href{http://arxiv.org/abs/1612.07698}{{\tt arXiv:1612.07698}}].

\bibitem{Boudaud:2018oya}
M.~Boudaud, T.~Lacroix, M.~Stref, and J.~Lavalle, {\it {Robust cosmic-ray constraints on $p$-wave annihilating MeV dark matter}},  {\em Phys. Rev. D} {\bf 99} (2019), no.~6 061302, [\href{http://arxiv.org/abs/1810.01680}{{\tt arXiv:1810.01680}}].

\bibitem{Arguelles:2019ouk}
C.~A. Arg\"uelles, A.~Diaz, A.~Kheirandish, A.~Olivares-Del-Campo, I.~Safa, and A.~C. Vincent, {\it {Dark matter annihilation to neutrinos}},  {\em Rev. Mod. Phys.} {\bf 93} (2021), no.~3 035007, [\href{http://arxiv.org/abs/1912.09486}{{\tt arXiv:1912.09486}}].

\bibitem{Planck:2018vyg}
{\bf Planck} Collaboration, N.~Aghanim et~al., {\it {Planck 2018 results. VI. Cosmological parameters}},  {\em Astron. Astrophys.} {\bf 641} (2020) A6, [\href{http://arxiv.org/abs/1807.06209}{{\tt arXiv:1807.06209}}]. [Erratum: Astron.Astrophys. 652, C4 (2021)].

\bibitem{SimonsObservatory:2018koc}
{\bf Simons Observatory} Collaboration, P.~Ade et~al., {\it {The Simons Observatory: Science goals and forecasts}},  {\em JCAP} {\bf 02} (2019) 056, [\href{http://arxiv.org/abs/1808.07445}{{\tt arXiv:1808.07445}}].

\bibitem{Abazajian:2019eic}
K.~Abazajian et~al., {\it {CMB-S4 Science Case, Reference Design, and Project Plan}},  [\href{http://arxiv.org/abs/1907.04473}{{\tt arXiv:1907.04473}}].

\bibitem{Pospelov:2010hj}
M.~Pospelov and J.~Pradler, {\it {Big Bang Nucleosynthesis as a Probe of New Physics}},  {\em Ann. Rev. Nucl. Part. Sci.} {\bf 60} (2010) 539--568, [\href{http://arxiv.org/abs/1011.1054}{{\tt arXiv:1011.1054}}].

\bibitem{Hochberg:2014dra}
Y.~Hochberg, E.~Kuflik, T.~Volansky, and J.~G. Wacker, {\it {Mechanism for Thermal Relic Dark Matter of Strongly Interacting Massive Particles}},  {\em Phys. Rev. Lett.} {\bf 113} (2014) 171301, [\href{http://arxiv.org/abs/1402.5143}{{\tt arXiv:1402.5143}}].

\bibitem{Hochberg:2014kqa}
Y.~Hochberg, E.~Kuflik, H.~Murayama, T.~Volansky, and J.~G. Wacker, {\it {Model for Thermal Relic Dark Matter of Strongly Interacting Massive Particles}},  {\em Phys. Rev. Lett.} {\bf 115} (2015), no.~2 021301, [\href{http://arxiv.org/abs/1411.3727}{{\tt arXiv:1411.3727}}].

\bibitem{Kulkarni:2022bvh}
S.~Kulkarni, A.~Maas, S.~Mee, M.~Nikolic, J.~Pradler, and F.~Zierler, {\it {Low-energy effective description of dark $Sp(4)$ theories}},  {\em SciPost Phys.} {\bf 14} (2023), no.~3 044, [\href{http://arxiv.org/abs/2202.05191}{{\tt arXiv:2202.05191}}].

\bibitem{Witten:1983tw}
E.~Witten, {\it {Global Aspects of Current Algebra}},  {\em Nucl. Phys. B} {\bf 223} (1983) 422--432.

\end{thebibliography}\endgroup

\end{document}